\documentclass[referee]{raa}            

\usepackage{graphicx,times}             
\usepackage{natbib}
\usepackage{amssymb,amsmath}
\bibpunct{(}{)}{;}{a}{}{,}

\usepackage[a4paper=true,dvipdfm=true,pagebackref=true]{hyperref}
\hypersetup{colorlinks = true, linkcolor = green, anchorcolor = red, citecolor = blue, filecolor = red, pagecolor = red, urlcolor = red}
\usepackage{longtable}
\usepackage{booktabs}

\begin{document}

   \title{$BV(RI)_{c}$ photometric study of three variable PMS stars in the field of V733 Cephei}

   \volnopage{Vol.0 (20xx) No.0, 000--000}      
   \setcounter{page}{1}          

   \author{Sunay Ibryamov
      \inst{1}
   \and Evgeni Semkov
      \inst{2}
   \and Stoyanka Peneva
      \inst{2}
	 \and Kristina Gocheva
      \inst{1}
   }

   \institute{Department of Physics and Astronomy, University of Shumen,
             115, Universitetska Str., 9700 Shumen, Bulgaria; {\it sibryamov@shu.bg}\\
        \and
             Institute of Astronomy and National Astronomical Observatory, Bulgarian Academy of Sciences, 72, Tsarigradsko Shose Blvd., 1784 Sofia, Bulgaria\\
\vs\no
   {\small Received~~20xx month day; accepted~~20xx~~month day}}

\abstract{
This paper report results from the first long-term $BV(RI)_{c}$ photometric CCD observations of three variable pre-main-sequence stars collected during the period from February 2007 to January 2020.
The investigated stars are located in the field of the PMS star V733 Cep within the Cepheus OB3 association.
All stars from our study show rapid photometric variability in all-optical passbands.
In this paper, we describe and discuss the photometric behavior of these stars and the possible reasons for their variability.
In the light variation of two of the stars we found periodicity.
}

   \authorrunning{S. Ibryamov, E. Semkov, S. Peneva \& K. Gocheva}            
   \titlerunning{$BV(RI)_{c}$ photometric study of three variable PMS stars in the field of V733 Cephei}  

   \maketitle

%
%
\section{Introduction}           
\label{sect:intro}

Pre-main-sequence (PMS) stars are important in the study of star formation.
They are divided into two main types $-$ the low mass (M $\leq$ 2M$_{\odot}$) T Tauri stars (TTS) and the more massive (2M$_{\odot}$ $\leq$ M $\leq$ 8M$_{\odot}$) Herbig Ae/Be stars (HAEBES).
The first systematic study of TTS as a separate class variable was made by \cite{Joy+1945}, and their characteristics were reviewed by \cite{Petrov+2003} and \cite{Cram+etal+1989}.
TTS are dwarf stars with photospheric spectra ranging from late F to M (\citealt{Petrov+2003}).
These stars show rapid irregular light variations and emission spectra, and they are associated with molecular clouds, dark and bright nebulae.

\newpage
TTS are separated into two subgroups $-$ classical T Tauri stars (CTTS), surrounded by a massive accreting circumstellar disk, and weak-line T Tauri stars (WTTS), which show no signs of disk accretion (\citealt{Menard+Bertout+1999}).
Variable mass accretion rate and the presence of hot and cool spots on the stellar surface are possible reasons for the variability of CTTS.
The variability of WTTS is often associated with the presence of cool spots or groups of spots on the stellar surface and flare-like events (\citealt{Herbst+etal+2007}).

Some of the PMS stars undergo obscuration in which their brightness decrease with an amplitude of up to 2-3 mag.
Such decreases in the brightness are observed in both CTTS and WTTS, and its clearest form is observed in young stars of an earlier spectral class (HAEBES) (\citealt{Petrov+2003}).
These PMS stars are called UXors, named after their prototype UX Orionis.
The observed deep Algol-type minima in the light curves of UXors probably are caused by obscuration of the star from circumstellar dust clouds with different density or edge-on circumstellar disks.
These ideas were discussed by \cite{Grinin+1988}, \cite{Voshchinnikov+1989}, \cite{Grinin+etal+1991}, \cite{Dullemond+etal+2003}, and other authors.
In the deep minima, the color indices of UXors often become bluer.
This specific color variability is known as the blueing effect (see \citealt{Bibo+The+1990}).
The star's color initially becomes redder (when its light is covered by dust clumps or filaments), but in a large extinction, the scattered light from the dust clouds begins to dominate, and the star becomes bluer.
Illustrative figures for the blueing effect in UXors can be seen in the works of \cite{Herbst+Shevchenko+1999} (Figs. 6-8 therein) and \cite{Semkov+etal+2015} (Fig. 3 therein).

The stars included in the present study are located near to the well-studied PMS star V733 Cep (see \citealt{Reipurth+etal+2007}, \citealt{Semkov+Peneva+2008}, \citealt{Peneva+etal+2010}) within the Cep OB3 association.
During its photometric study of V733 Cep \cite{Semkov+Peneva+2008} discovered the variability of three nearby stars.
For short these stars were named Var. 1, Var. 2, and Var. 3, corresponding to 2MASS J22534654+6234582, 2MASS J22533629+6231446, and 2MASS J22531578+6235262, respectively.
\cite{Munari+2009} estimated their brightness on 40 plates from the archive of the Asiago 67/92-cm Schmidt telescope distributed over the period from August 1971 to November 1978.

In this paper we present the results from the multicolor observations of these stars that cover 13 years.
We aim to describe and discuss the optical photometric behavior of the investigated stars and the possible reasons for their variability.
This paper is organized as follows.
In Section 2, we give information about the telescopes and cameras used to perform the observations.
In Section 3, we describe the results obtained for the investigated stars and their interpretation.
In Concluding remarks, we briefly present the main results of our study.

\section{Observations and data reduction}
\label{sect:Obs}
We obtained $BV(RI)_{c}$ photometric CCD observations of the field of V733 Cep in the period from February 2007 to January 2020.
The observations were performed with four telescopes: the 2-m Ritchey-Chr\'{e}tien-Coud\'{e} (RCC), the 50/70-cm Schmidt and the 60-cm Cassegrain telescopes administered by the Rozhen National Astronomical Observatory in Bulgaria, and the 1.3-m Ritchey-Chr\'{e}tien (RC) telescope administered by the Skinakas Observatory\footnote{Skinakas Observatory is a collaborative project of the University of Crete, the Foundation for Research and Technology, Greece, and the Max-Planck-Institut f{\"u}r Extraterrestrische Physik, Germany.} of the University of Crete in Greece.

Eight different CCD cameras were used to obtain the observations, as follows: VersArray 1300B and Andor iKon-L at the 2-m RCC telescope; Photometrics CH360 and Andor DZ436-BV at the 1.3-m RC telescope; SBIG ST-8, SBIG STL-11000M and FLI PL16803 at the 50/70-cm Schmidt telescope; and FLI PL09000 at the 60-cm Cassegrain telescope.
The technical parameters and chip specifications for the cameras used are given in \cite{Ibryamov+etal+2015}.
Because the camera Andor iKon-L at the 2-m RCC telescope was installed in 2018, its specifications are given in \cite{Ibryamov+Semkov+2020}.

All frames were taken through a standard Johnson-Cousins ($BVR_{c}I_{c}$) set of filters.
The data were reduced using an \textsc{idl} based \textsc{daophot} subroutine.
All data were analyzed using the same aperture, which was chosen to have a 3 arcsec radius (the background annulus was taken from 10 arcsec to 13 arcsec).
As a reference, the improved $BV(RI)_{c}$ comparison sequence in the field around V733 Cep reported in \cite{Peneva+etal+2010} was used.
The calibrations of the comparison stars were made with the 1.3-m RC telescope during six clear nights (four in 2007 and two in 2009), as the standard stars from \cite{Landolt+1992} were used as a reference.
The finding chart of the comparison sequence and the coordinates of the stars can be found in \cite{Semkov+Peneva+2008}.

\section{Results and discussion}
\label{sect:Res}

\begin{figure}[h!]
\begin{center}
\includegraphics[width=9cm]{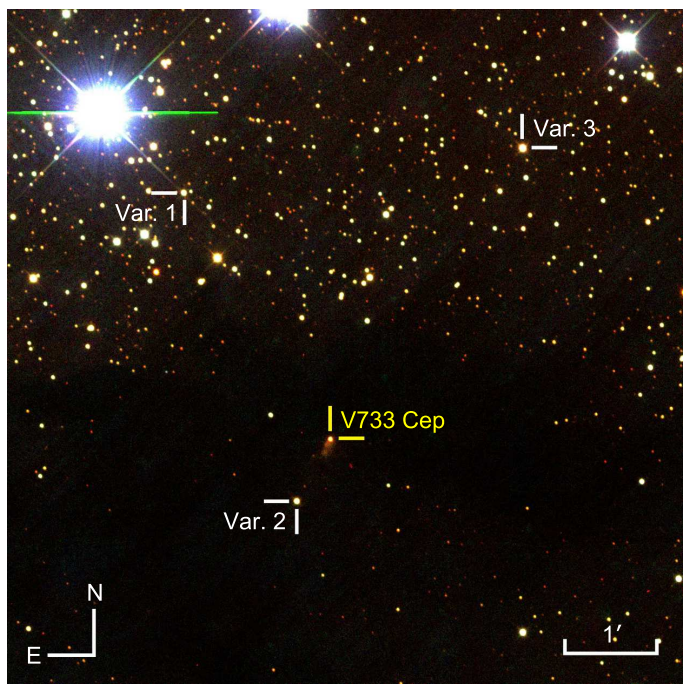}
\caption{A three-color image of the field around V733 Cep obtained on 12 August 2015 with the 1.3-m RC telescope at the Skinakas Observatory. The stars from our study and V733 Cep are marked.}\label{Fig:field}
\end{center}
\end{figure}

Figure~\ref{Fig:field} shows a three-color image of the field around V733 Cep, where the positions of the investigated stars are marked.
These stars are probable members of the Cep OB3 association.
The \textit{Gaia} mean distances (\citealt{Bailer-Jones+etal+2018}) to the three stars are as follows: 817 pc to Var. 1, 725 pc to Var. 2, and 826 pc to Var. 3.
The distance to the Cep OB3 association was determined as 725 pc by \cite{Blaauw+etal+1959} and as 900$\pm$100 pc by \cite{Moreno-Corral+etal+1993}.

The results from our long-term observations of the stars are summarized in Table~\ref{Tab:photV1}, Table~\ref{Tab:photV2}, and Table~\ref{Tab:photV3} in the Appendix\footnote{The tables with the photometric data will be available also via CDS VizieR Online Data Catalogue.}.
The columns contain Date (dd.mm.yyyy format) and Julian date (JD) of the observations, the measured magnitudes of the star, and the telescopes and CCD cameras used.
The average value of the errors in the reported magnitudes is 0.01-0.02 mag for the data in $I_{c}$ and $R_{c}$ bands, and 0.02-0.05 mag for the data in $V$ and $B$ bands.
Table~\ref{Tab:comparison} shows the mean magnitudes of the investigated stars from \cite{Munari+2009} and the present study.

\begin{table}[h!]
  \caption{Comparisons of the mean magnitudes of Var. 1, Var. 2, and Var. 3.}\label{Tab:comparison}
  \begin{center}
  \begin{tabular}{l|cccc|cccc}
	  \hline\hline
	  \noalign{\smallskip}
Star  & $\overline{I_{c}}$ & $\overline{R_{c}}$ & $\overline{V}$ & $\overline{B}$ & $\overline{I_{c}}$ & $\overline{R_{c}}$ & $\overline{V}$ & $\overline{B}$ \\
      & \multicolumn{4}{c}{(\citealt{Munari+2009})}                           & \multicolumn{4}{c}{(our study)}\\
   \noalign{\smallskip}
   \hline
   \noalign{\smallskip}
Var. 1& 14.49 & - & 16.93 & 18.80 & 14.91 & 16.26 & 17.58 & 19.56 \\
Var. 2& 13.83 & - & 16.40 & 17.68 & 13.97 & 15.26 & 16.51 & 18.48 \\
Var. 3& 13.01 & - & 15.03 & 16.95 & 13.25 & 14.37 & 15.50 & 17.36 \\
   	  \hline \hline
  \end{tabular}
  \end{center}
	\end{table}

\subsection{2MASS J22534654$+$6234582 (Var. 1 hereafter)}

\cite{Semkov+Peneva+2008} reported the observed values of the brightness of Var. 1 for the period from February 2007 to February 2008.
According to the authors, the star shows a very high amplitude of brightness variation, and it is probably a long-period variable of Mira-type.
\cite{Munari+2009} measured the star's brightness for the period 1971 - 1978 and reported that its variability at $I_{c}$ is much less pronounced than listed by \cite{Semkov+Peneva+2008}.
The author discussed the characteristics that should have a Mira star, which lies within the Galaxy in that direction with $\overline{V}$=16.9 mag.
\cite{Munari+2009} concluded that $\Delta$$I_{c}$=2.36 mag reported by \cite{Semkov+Peneva+2008} but not confirmed by their data would be far too large for such a Mira.
The star was included in the list of young stellar objects: class II published by \cite{Allen+etal+2012}.

\begin{figure}[h!]
\begin{center}
\includegraphics[width=\textwidth]{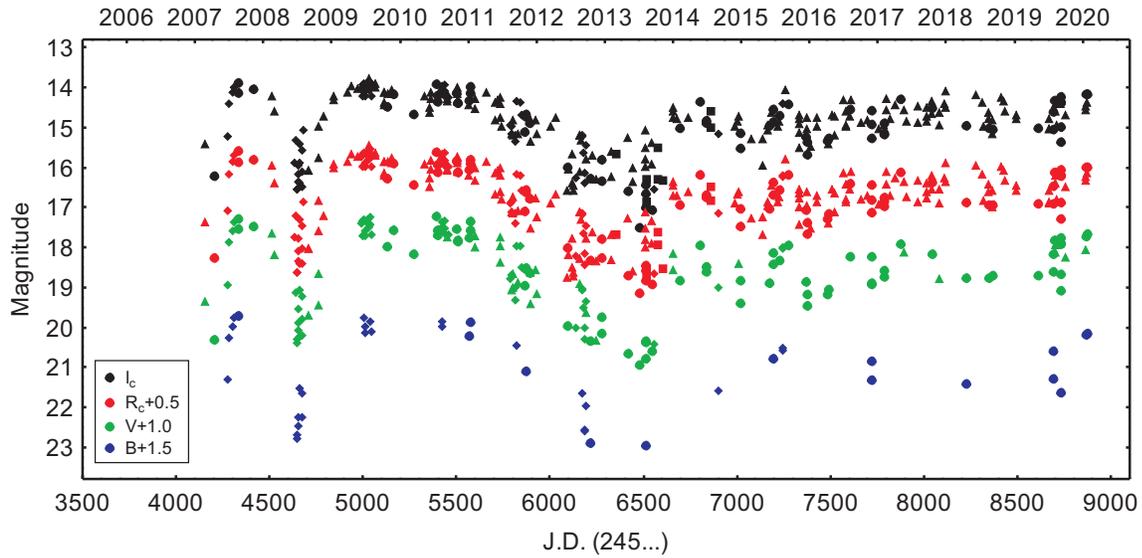}
\caption{$BV(RI)_{c}$ light curves of Var. 1 for the period February 2007 - January 2020. The circles denote the photometric data from the 2-m RCC telescope, the diamonds represent the photometric data from the 1.3-m RC telescope, the triangles mark the photometric data from the 50/70-cm Schmidt telescope, and the squares denote the photometric data from the 60-cm Cassegrain telescope. In very low light, the brightness of the star, especially in the $V$ and $B$ bands is under the photometric limit of the telescopes used.}\label{Fig:curvesV1}
\end{center}
\end{figure}

The $BV(RI)_{c}$ light curves of Var. 1 from our observations are shown in Fig.~\ref{Fig:curvesV1}.
During our monitoring, the star shows rapid variability and fading events in the light curves.
As can be seen in Fig.~\ref{Fig:curvesV1} the star's brightness varies in wide ranges: 13.77-17.49 mag in the $I_{c}$ band, 14.93-18.62 mag in the $R_{c}$ band, 16.20-19.92 mag in the $V$ band, and 18.19-21.45 mag in the $B$ band.

In the period February 2007 - January 2020, several well defined deep minima in the light curves of Var. 1 are registered in all-bands.
They have different amplitude, shape, and duration.
The first minimum is registered in April 2007, at the beginning of our photometric monitoring of the star.
The second well defined minimum is registered in June - July 2008.
The third minimum is longer than the others, and it is registered in the period June 2012 - November 2013.
During this minimum, two deeper drops in the brightness of the star are also observed.
The deepest dip ($\Delta$$I_{c}$=3.72 mag) we registered in the second half of 2013.
Out of deep minima, the star shows significant rapid brightness variation in the time scale of days and weeks.

\begin{figure}[h!]
\begin{center}
\includegraphics[width=8cm]{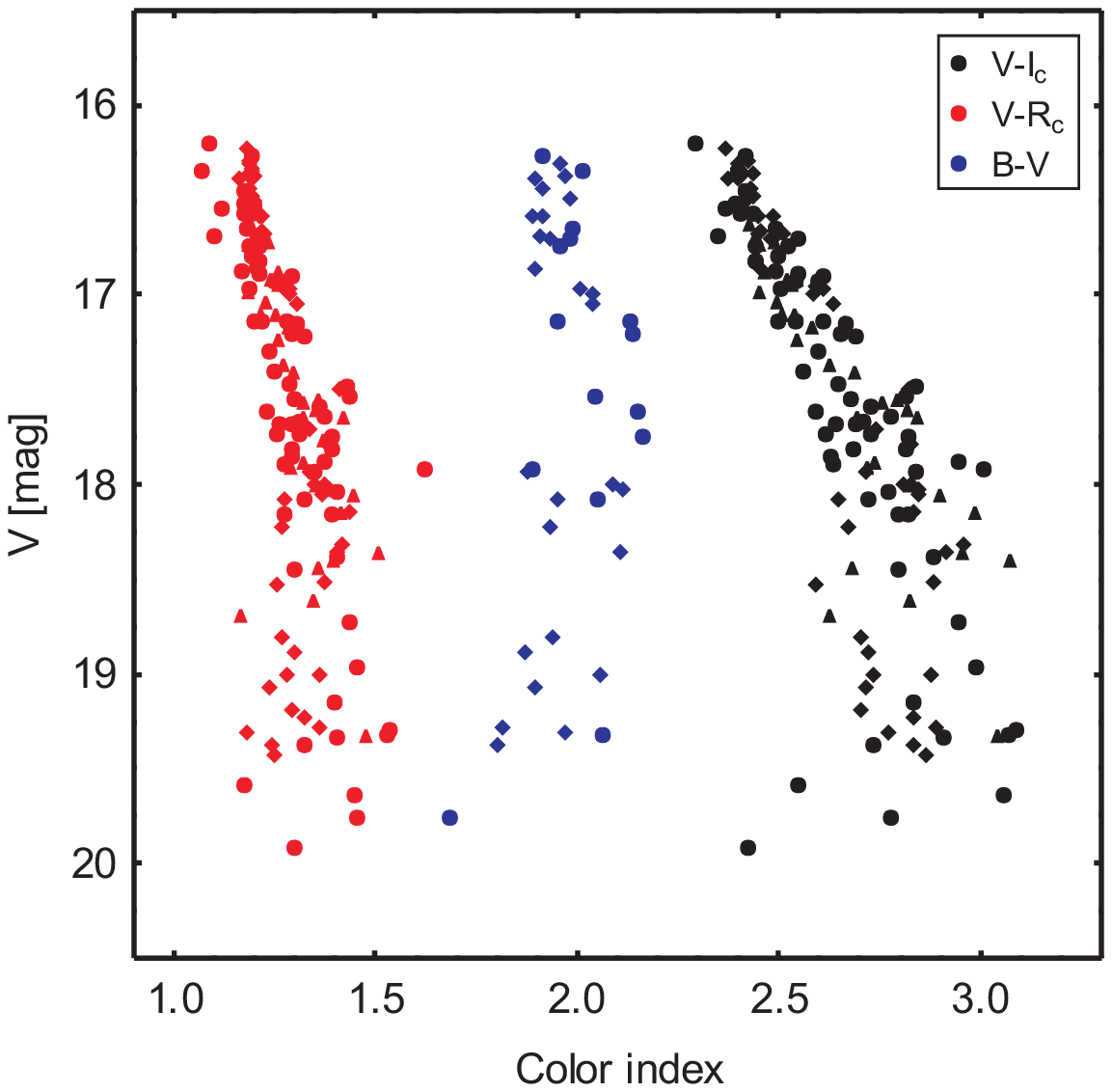}
\caption{Color indices versus the $V$ magnitude of Var. 1. The symbols used are as in Fig.~\ref{Fig:curvesV1}.}\label{Fig:colorV1}
\end{center}
\end{figure}

An important result from our study of Var. 1 is the registered change in its color at the deep minima.
We constructed three diagrams of color - magnitude, which are displayed in Fig.~\ref{Fig:colorV1}.
It is seen that the star becomes redder as it fades, and during the deep minima real blueing effect is observed on all diagrams.
Most probably, the observed color reverse is caused by the scattered light from small dust grains.
In this case, it can be assumed that the ratio of scattered light to direct light increases and the star's color turns bluer.

\newpage
Our analysis of the collected photometric data of Var. 1 shows that the observed large amplitude minima in the brightness and the blueing effect in the color-magnitude diagrams are independent evidence of its UXor-type variability.
The different shapes and amplitudes of the minima in the star's light curves give grounds to predict a variety of obscuration reasons: massive dust clumps, the precession and inhomogeneous structures of the circumstellar disk,  planetesimals at different stages of formation, and clouds of protostellar material orbiting at the vicinity of the star.

\begin{figure}[h!]
\begin{center}
\includegraphics[width=7cm]{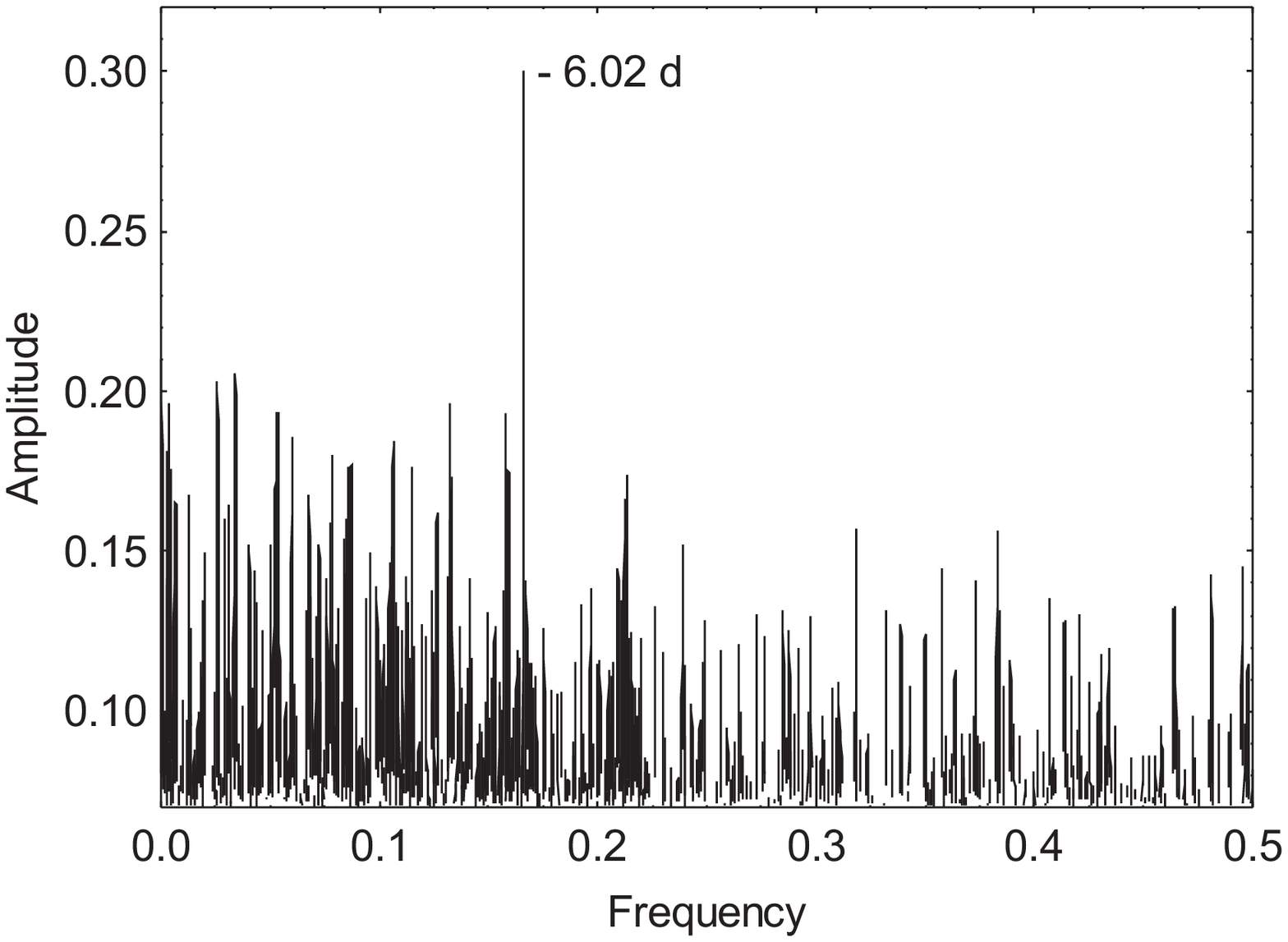}
\includegraphics[width=7cm]{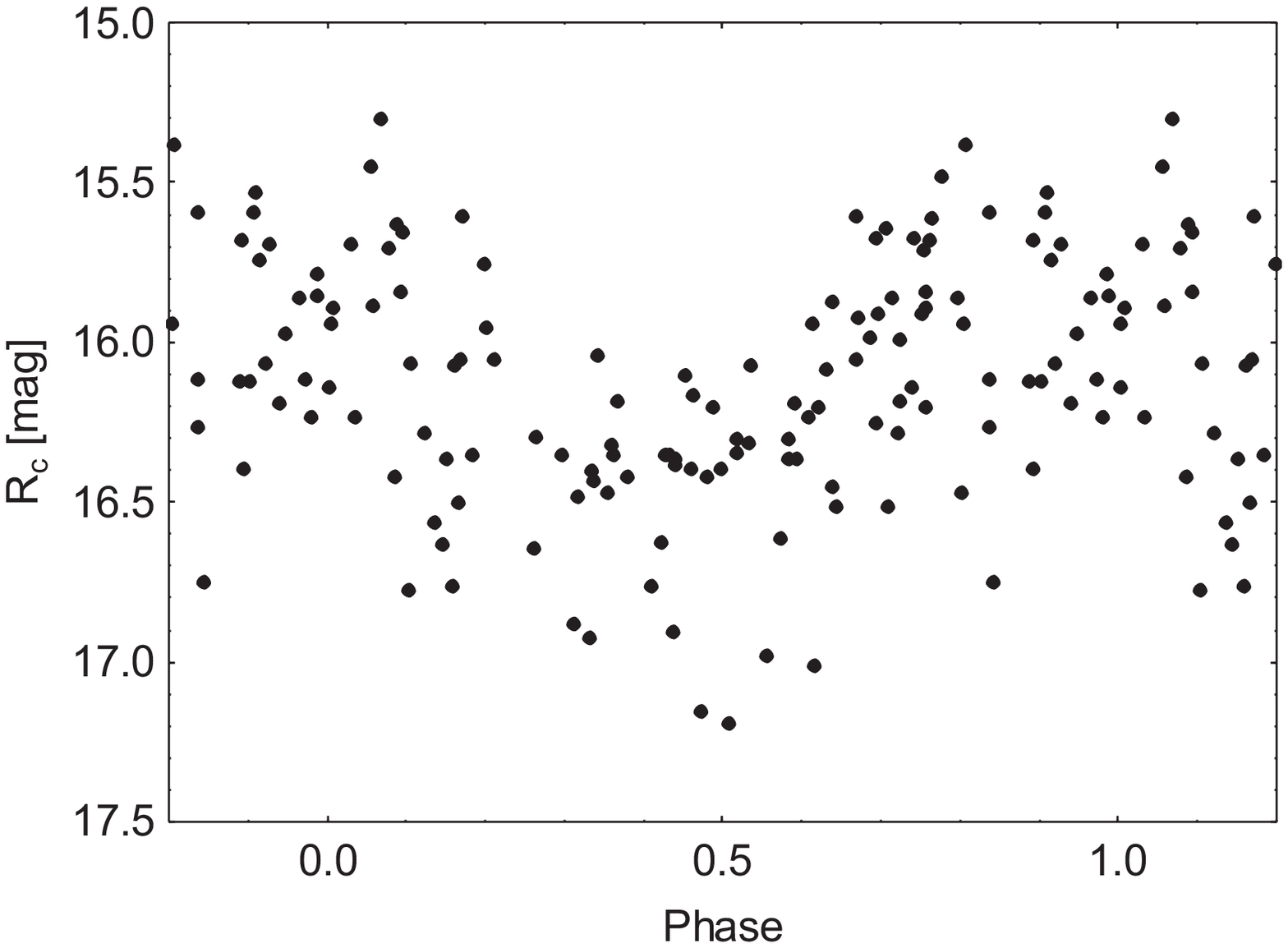}
\caption{Left: Periodogram analysis of the photometric data of Var. 1. Right: $R_{c}$ band phased light curve of the star.}\label{Fig:periodV1}
\end{center}
\end{figure}

Another important result from our study of Var. 1 is the identification of previously unknown periodicity in its light variation.
We carried out a periodicity search in the photometric behavior of the star by \textsc{period04} (\citealt{Lenz+Breger+2005}).
Initially, we used all our data to search for periodicity, but we did not find any reliable one.
Using the data received after December 2013, when the star's brightness varies around some intermediate level in maximal light, our time-series analysis indicates a 6.02$^{d}$ period.
The periodogram analysis of our data of the star and its $R_{c}$ band phased light curve according to the found period is plotted in Fig.~\ref{Fig:periodV1}.
The period remained stable during the time interval of six years and it is a typical rotational period for a young stellar object.
Such periodicity could be caused by rotation modulation of cool spots on the stellar surface.
According to \cite{Herbst+etal+1994}, cool spots may last for hundreds or thousands of rotation of the young star, as in the case of Var. 1.

\subsection{2MASS J22533629$+$6231446 (Var. 2 hereafter)}

Var. 2 is located at 45 arcsec from V733 Cep and it is embedded at LDN 1216.
According to \cite{Semkov+Peneva+2008} the star is probably a PMS object.
\cite{Munari+2009} reported that Var. 2 does not vary on the 1971 - 1978 plates.
The star was included in the list of young stellar objects: class II published by \cite{Allen+etal+2012}.

\begin{figure}[h!]
\begin{center}
\includegraphics[width=\textwidth]{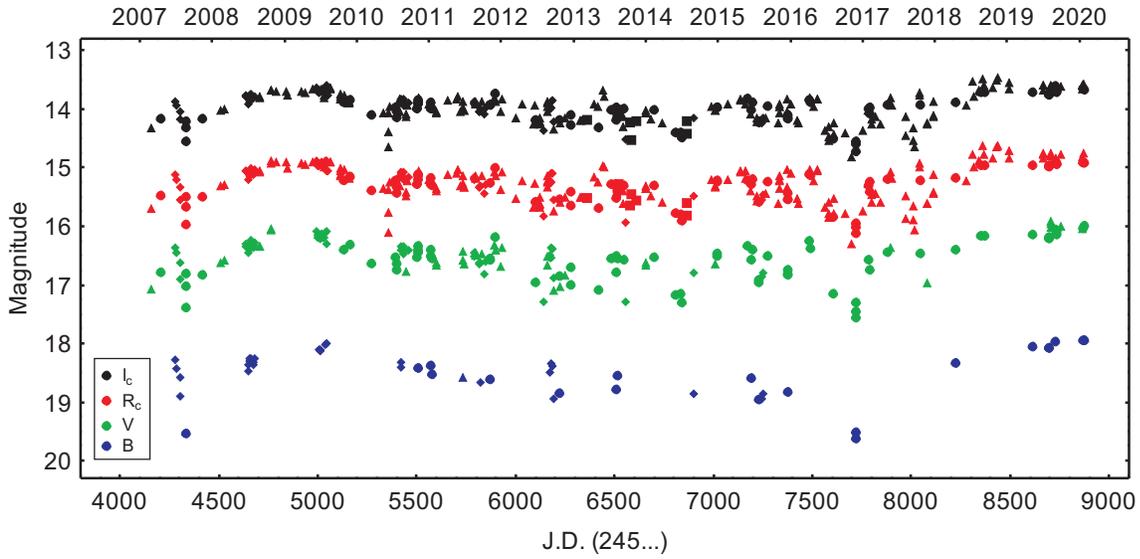}
\caption{$BV(RI)_{c}$ light curves of Var. 2 for the period February 2007 - January 2020. The symbols used are as in Fig.~\ref{Fig:curvesV1}.}\label{Fig:curvesV2}
\end{center}
\end{figure}

\newpage
The $BV(RI)_{c}$ light curves of Var. 2 from our monitoring are shown in Fig.~\ref{Fig:curvesV2}.
The available data suggest that during our observations the star exhibits strong irregular variability in all bands.
Short-time increases and decreases in the star's brightness are seen during the same period.
The reasons for the observed variability of Var. 2 may be different: variable accretion rate, the presence of hot and cool spots on the stellar surface, irregular obscuration by circumstellar material.
Its brightness variation during the whole period of our observations is in the range 13.46-14.82 mag in the $I_{c}$ band, 14.63-16.30 mag in the $R_{c}$ band, 15.92-17.54 mag in the $V$ band, and 17.94-19.60 mag in the $B$ band.
By comparing our mean values of Var. 2 (Table~\ref{Tab:comparison}) with those of \cite{Munari+2009}, we get a good match for the $I_{c}$ and $V$ bands and difference of 0.8 mag for the $B$ band.

\begin{figure}[h!]
\begin{center}
\includegraphics[width=8cm]{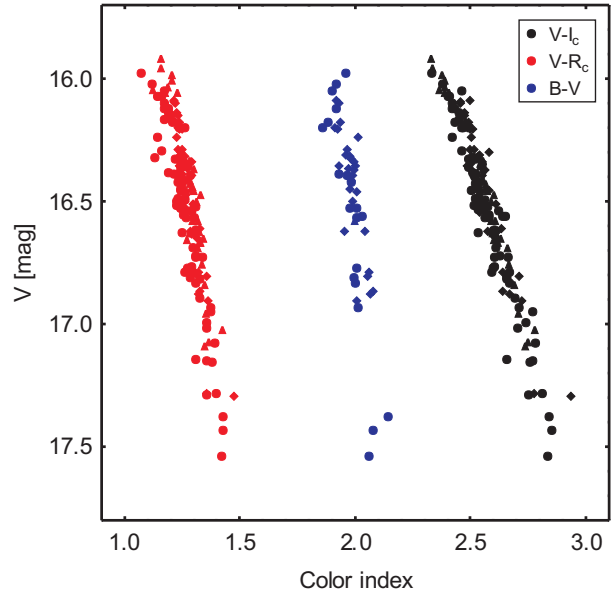}
\caption{Color indices versus the $V$ magnitude of Var. 2. The symbols used are as in Fig.~\ref{Fig:curvesV1}.}\label{Fig:colorV2}
\end{center}
\end{figure}

The measured color indices versus the $V$ magnitude of Var. 2 are plotted on Fig.~\ref{Fig:colorV2}.
As can be seen, the star becomes redder as it fades and the color reverse is not observed.
Such color variation is typical for both CTTS and WTTS with the presence of spots or groups of spots on the stellar surface.
In case that the decreases in the brightness of Var. 2 are caused by obscuration, the registered amplitudes are too small to show indication for the color reverse.
The direct light of the star is not enough suppressed to allow the scattered component to emerge.
Evidence of periodicity in the brightness variability of Var. 2 is not detected.

\subsection{2MASS J22531578+6235262 (Var. 3 hereafter)} 

\cite{Munari+2009} reported that Var. 3 does not vary on the 1971 - 1978 plates.
The star was included in the list of young stellar objects: class III published by \cite{Allen+etal+2012}.

\begin{figure}[h!!!]
\begin{center}
\includegraphics[width=\textwidth]{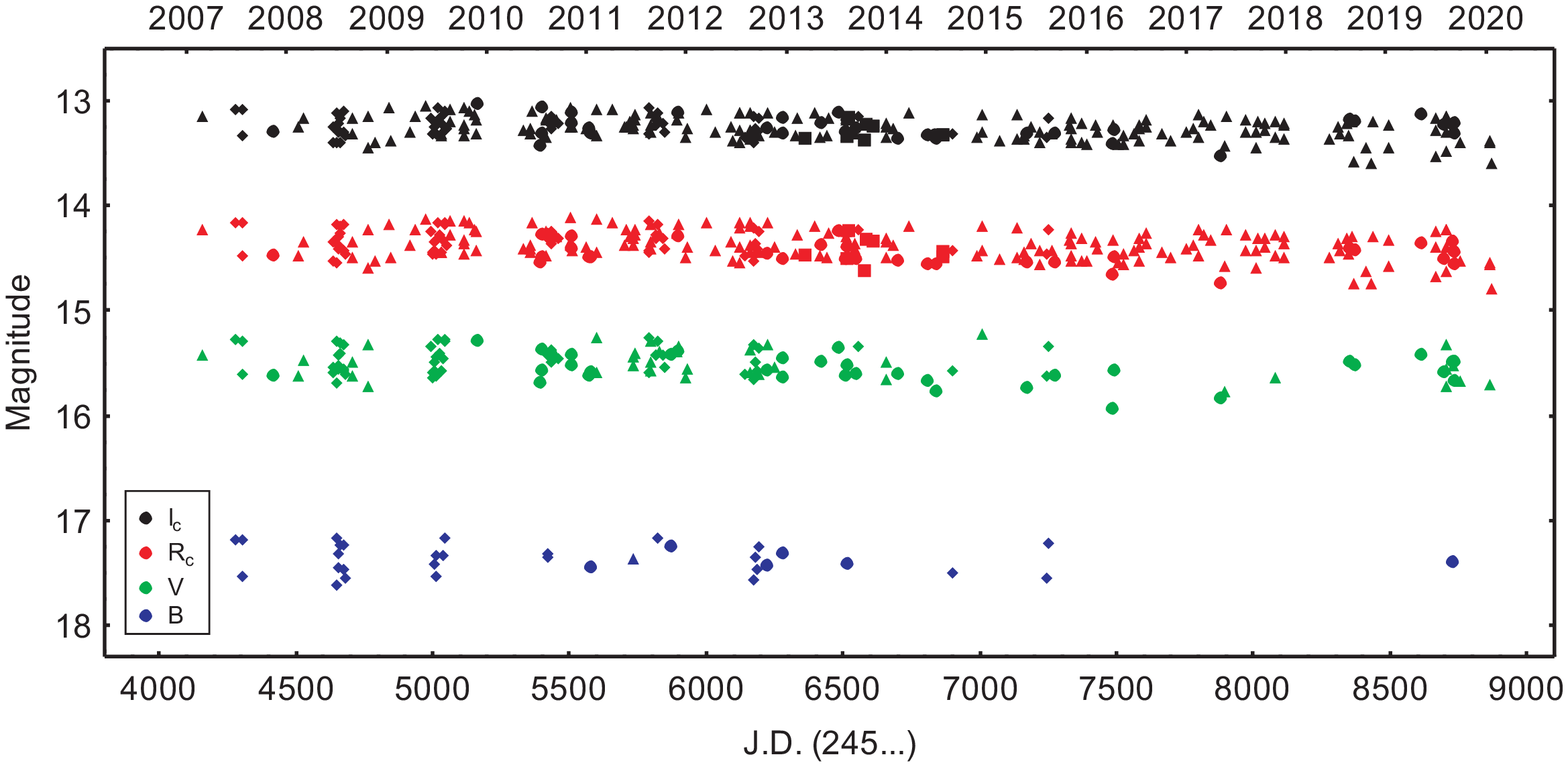}
\caption{$BV(RI)_{c}$ light curves of Var. 3 for the period February 2007 - January 2020. The symbols used are as in Fig.~\ref{Fig:curvesV1}.}\label{Fig:curvesV3}
\end{center}
\end{figure}

\newpage
The $BV(RI)_{c}$ light curves of the star from our observations are shown in Fig.~\ref{Fig:curvesV3}.
As can be seen that during the period of our monitoring Var. 3 shows a light variation with small amplitudes in all bands.
The observed amplitudes of its variability are $\Delta$$I_{c}$=0.57 mag, $\Delta$$R_{c}$=0.67 mag, and $\Delta$$V$=0.70 mag.
Usually, such low amplitude variability is typical for low-mass WTTS and it is caused by spots on the stellar surface.

\newpage
We found that the total brightness of Var. 3 gradually decreases during the whole period of our observations.
Using a linear approximation for all our data, we calculated the following values for the rate of decreases: 0.92$\times$10$^{-2}$mag $yr^{-1}$ for the $I_{c}$ band, 1.23$\times$10$^{-2}$mag $yr^{-1}$ for the $R_{c}$ band, and 1.13$\times$10$^{-2}$mag $yr^{-1}$ for the $V$ band.
By comparing our mean values of the star (Table~\ref{Tab:comparison}) with those of \cite{Munari+2009}, we get a significant difference in all bands.
This is an expected result given that the total star's brightness decreases over time.

\begin{figure}[h!]
\begin{center}
\includegraphics[width=8cm]{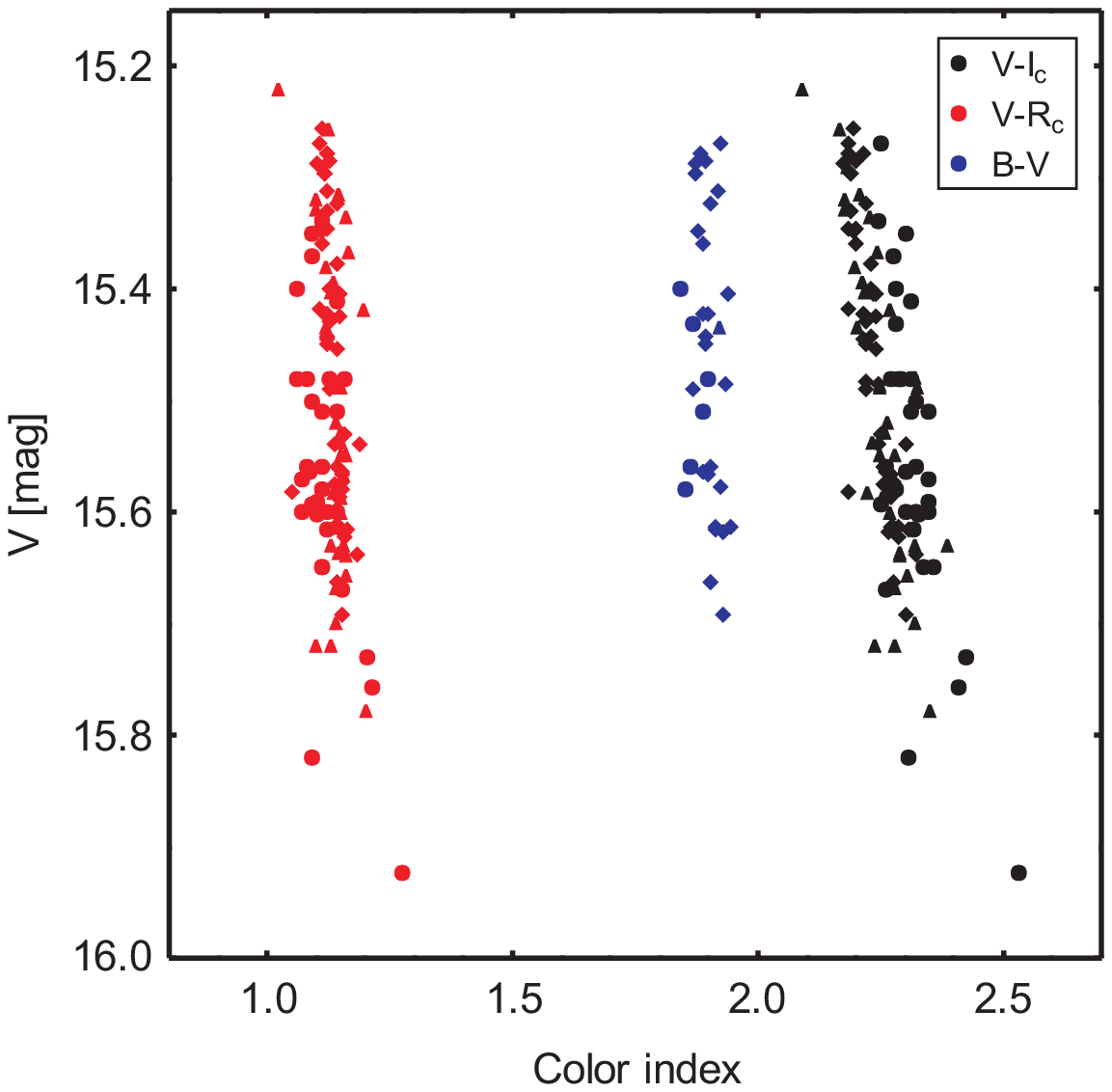}
\caption{Color indices versus the $V$ magnitude of Var. 3. The symbols used are as in Fig.~\ref{Fig:curvesV1}.}\label{Fig:colorV3}
\end{center}
\end{figure}

\begin{figure}[h!]
\begin{center}
\includegraphics[width=7cm]{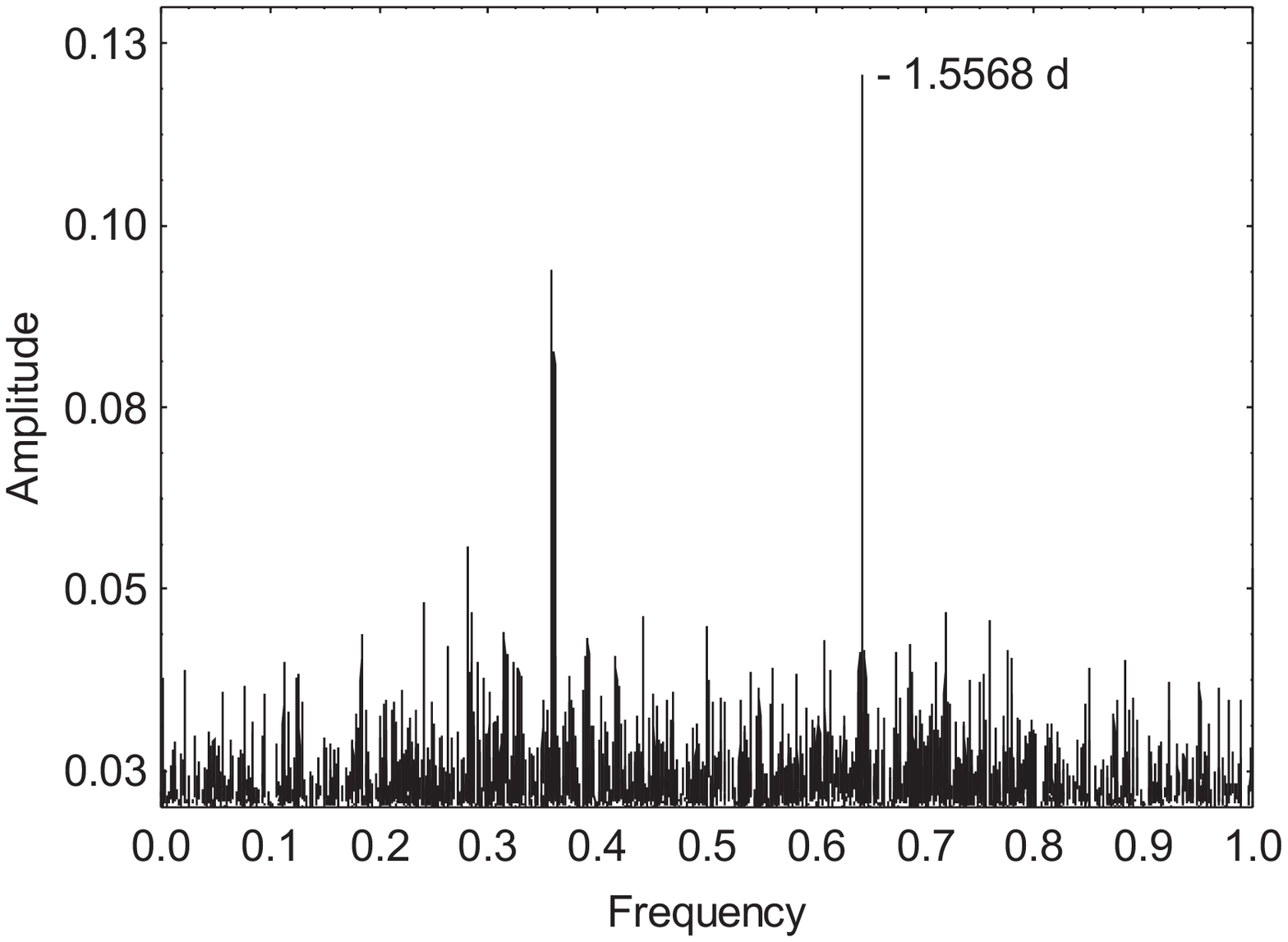}
\includegraphics[width=7cm]{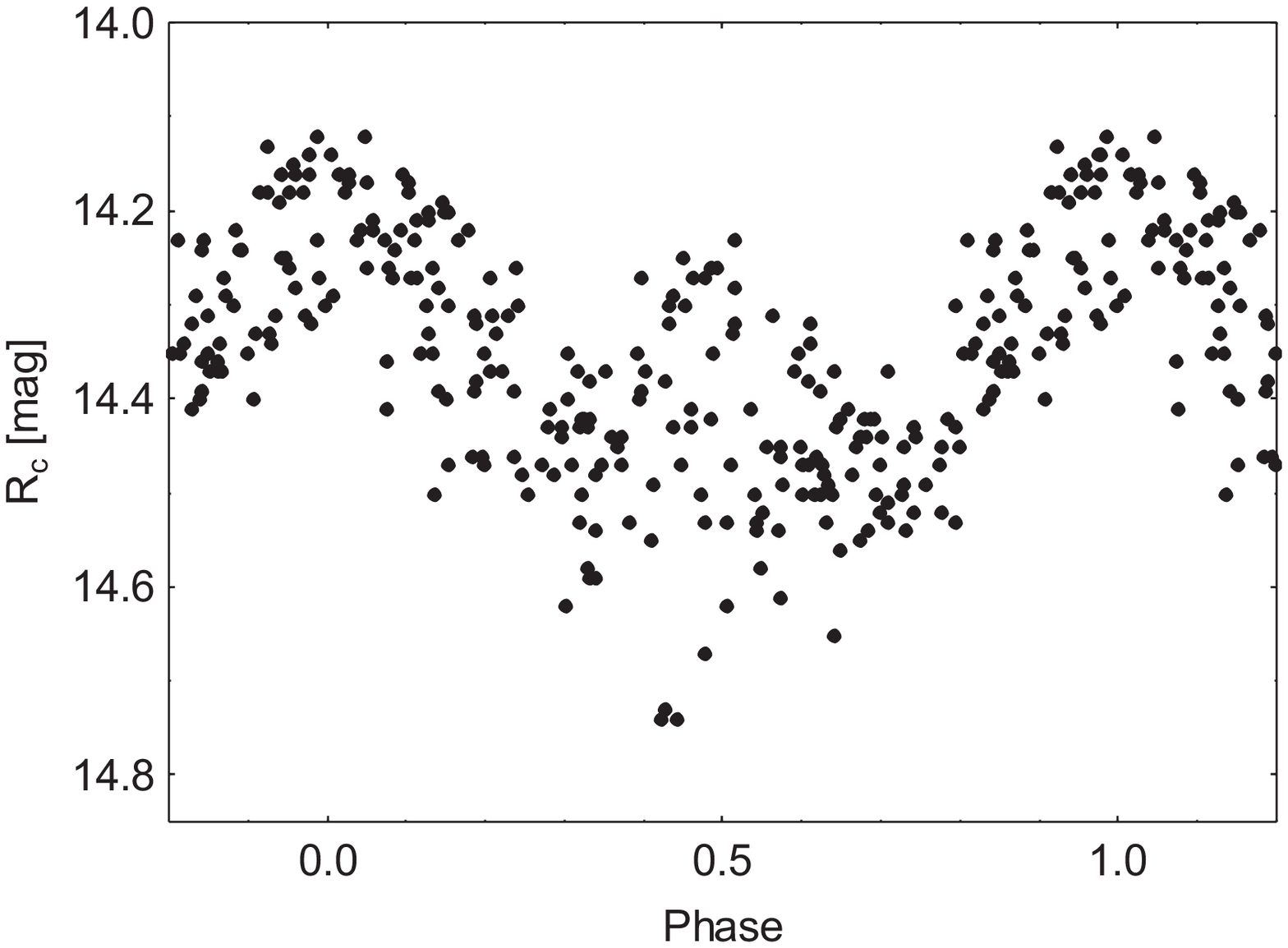}
\caption{Left: Periodogram analysis of the photometric data of Var. 3. Right: $R_{c}$ band phased light curve of the star.}\label{Fig:periodV3}
\end{center}
\end{figure}

The measured color indices versus the $V$ magnitude of Var. 3 are plotted in Fig.~\ref{Fig:colorV3}.
The registered color variation is typical of WTTS, stars with cool spots, whose variability is produced by rotation of the spotted surface.
Our time-series analysis of the data of the star indicated a period of 1.5568$^{d}$.
This result confirms $P$=1.5565$^{d}$ given in the ASAS-SN Variable Stars Database (\citealt{Jayasinghe+etal+2018}).
The periodogram analysis of the data of Var. 3 and its $R_{c}$ band phased light curve are plotted in Fig.~\ref{Fig:periodV3}.
The period remained stable during the whole period of our observations and it is a typical rotational period for a WTTS.

\section{Concluding remarks}
\label{sect:conclusion remarks}

We presented and discussed the optical photometric behavior of three variable PMS stars in the field of the star V733 Cep.
Our observations cover 13 years (2007 - 2020) and represent the first long-term CCD photometric monitoring of the investigated objects.
The obtained results of our study can be summarized as follows:
(i) 2MASS J22534654$+$6234582 (Var. 1) is a young stellar object and it shows light and color variations inherent to UXor-type variability.
We found a 6.02$^{d}$ rotational periodicity in its light variation.
(ii) 2MASS J22533629$+$6231446 (Var. 2) shows the photometric characteristics of both CTTS and WTTS.
For its exact classification, spectral observations are necessary.
(iii) 2MASS J22531578+6235262 (Var. 3) is most probably a WTTS.
We confirmed its periodicity given in the ASAS-SN Variable Stars Database.

\begin{acknowledgements}
This work was partly supported by the Bulgarian Ministry of Education and Science under the National Program for Research ''Young Scientists and Postdoctoral Students''.
The authors thank the Director of Skinakas Observatory Prof. I. Papamastorakis and Prof. I. Papadakis for the award of telescope time.
ES and SP acknowledge partial support by grant DN 18-13/2017 from the Bulgarian National Science Fund.
This research has made use of NASA's Astrophysics Data System.
\end{acknowledgements}

\newpage
\appendix                  

\section{$BV(RI)_{c}$ photometric data of Var. 1, Var. 2, and Var. 3}

{\footnotesize
\begin{longtable}{cccccccc}
\caption{Photometric CCD observations of Var. 1.}\\
\hline\hline
\noalign{\smallskip}  
Date &	JD (245...) &	$I_{c}$ & $R_{c}$ & $V$ & $B$ & Tel & CCD\\
\noalign{\smallskip}  
\hline
\endfirsthead
\caption{Continued.}\\
\hline\hline
\noalign{\smallskip}  
Date &	JD (245...) &	$I_{c}$ & $R_{c}$ & $V$ & $B$ & Tel & CCD\\
\noalign{\smallskip}  
\hline
\noalign{\smallskip}  
\endhead
\hline
\label{Tab:photV1}
\endfoot
\noalign{\smallskip}
25.02.2007	&	4157.212	&	15.41	&	16.85	&	18.36	&	-	&	Sch	&	ST-8	\\
10.04.2007	&	4200.581	&	16.20	&	17.75	&	19.29	&	-	&	2-m	&	VA	\\
27.06.2007	&	4278.518	&	15.21	&	16.59	&	17.93	&	19.81	&	1.3-m	&	Phot	\\
04.07.2007	&	4285.523	&	14.40	&	15.66	&	16.86	&	18.75	&	1.3-m	&	Phot	\\
23.07.2007	&	4305.493	&	14.13	&	15.37	&	16.58	&	18.50	&	1.3-m	&	AND	\\
25.07.2007	&	4306.511	&	13.98	&	15.19	&	16.38	&	18.28	&	1.3-m	&	AND	\\
15.08.2007	&	4328.401	&	14.13	&	15.34	&	16.52	&	-	&	2-m	&	VA	\\
17.08.2007	&	4330.459	&	13.86	&	15.08	&	16.27	&	18.19	&	2-m	&	VA	\\
\hline \hline
\end{longtable}}

{\footnotesize
\begin{longtable}{cccccccc}
\caption{Photometric CCD observations of Var. 2.}\\
\hline\hline
\noalign{\smallskip}  
Date &	JD (245...) &	$I_{c}$ & $R_{c}$ & $V$ & $B$ & Tel & CCD\\
\noalign{\smallskip}  
\hline
\endfirsthead
\caption{Continued.}\\
\hline\hline
\noalign{\smallskip}  
Date &	JD (245...) &	$I_{c}$ & $R_{c}$ & $V$ & $B$ & Tel & CCD\\
\noalign{\smallskip}  
\hline
\noalign{\smallskip}  
\endhead
\hline
\label{Tab:photV2}
\endfoot
\noalign{\smallskip}
25.02.2007	&	4157.212	&	14.33	&	15.71	&	17.08	&	-	&	Sch	&	ST-8	\\
10.04.2007	&	4200.581	&	14.16	&	15.47	&	16.77	&	-	&	2-m	&	VA	\\
27.06.2007	&	4278.518	&	13.87	&	15.12	&	16.36	&	18.28	&	1.3-m	&	Phot	\\
04.07.2007	&	4285.523	&	13.93	&	15.19	&	16.45	&	18.43	&	1.3-m	&	Phot	\\
23.07.2007	&	4305.493	&	14.18	&	15.54	&	16.90	&	18.91	&	1.3-m	&	AND	\\
25.07.2007	&	4306.511	&	14.05	&	15.34	&	16.62	&	18.58	&	1.3-m	&	AND	\\
14.08.2007	&	4327.400	&	14.31	&	15.66	&	17.02	&	-	&	2-m	&	VA	\\
15.08.2007	&	4328.401	&	14.19	&	15.49	&	16.79	&	-	&	2-m	&	VA	\\
\hline \hline
\end{longtable}}

{\footnotesize
\begin{longtable}{cccccccc}
\caption{Photometric CCD observations of Var. 3.}\\
\hline\hline
\noalign{\smallskip}  
Date &	JD (245...) &	$I_{c}$ & $R_{c}$ & $V$ & $B$ & Tel & CCD\\
\noalign{\smallskip}  
\hline
\endfirsthead
\caption{Continued.}\\
\hline\hline
\noalign{\smallskip}  
Date &	JD (245...) &	$I_{c}$ & $R_{c}$ & $V$ & $B$ & Tel & CCD\\
\noalign{\smallskip}  
\hline
\noalign{\smallskip}  
\endhead
\hline
\label{Tab:photV3}
\endfoot
\noalign{\smallskip}
25.02.2007	&	4157.212	&	13.15	&	14.22	&	15.42	&	-	&	Sch	&	ST-8	\\
27.06.2007	&	4278.518	&	13.08	&	14.16	&	15.27	&	17.19	&	1.3-m	&	Phot	\\
23.07.2007	&	4305.493	&	13.34	&	14.48	&	15.61	&	17.53	&	1.3-m	&	AND	\\
25.07.2007	&	4306.511	&	13.08	&	14.16	&	15.29	&	17.18	&	1.3-m	&	AND	\\
06.11.2007	&	4411.214	&	13.28	&	14.46	&	15.60	&	-	&	2-m	&	VA	\\
12.02.2008	&	4509.229	&	13.24	&	14.47	&	15.63	&	-	&	Sch	&	STL-11	\\
29.02.2008	&	4526.300	&	13.16	&	14.35	&	15.48	&	-	&	Sch	&	STL-11	\\
15.06.2008	&	4632.578	&	13.40	&	14.53	&	15.58	&	-	&	1.3-m	&	AND	\\
\hline \hline
\end{longtable}}

\end{document}